
\input harvmac.tex
%
\def\figflag{I}

\noblackbox


%
\long\def\optional#1{}
\def\cmp#1{#1}         
%
%
\let\narrowequiv=\equiv
\def\equiv{\;\narrowequiv\;}

\def\tilde{\widetilde}
\fontdimen16\tensy=2.7pt\fontdimen17\tensy=2.7pt 

\def\ph{\varphi}
\def\ket#1{| #1 \rangle}         
\def\bra#1{\langle #1 |}         
\def\dd{\mskip 1.3mu{\rm d}\mskip .7mu} 
\global\newcount\figno \global\figno=1
\newwrite\ffile
\def\pfig#1#2{Fig.~\the\figno\pnfig#1{#2}}
\def\pnfig#1#2{\xdef#1{Fig. \the\figno}%
\ifnum\figno=1\immediate\openout\ffile=figs.tmp\fi%
\immediate\write\ffile{\noexpand\item{\noexpand#1\ }#2}%
\global\advance\figno by1}

%
%
\def\tfig#1{Fig.~\the\figno\xdef#1{Fig.~\the\figno}\global\advance\figno by1}
\def\ifig#1#2#3{\midinsert\vskip #3truein
    \narrower\narrower\noindent#1:#2\endinsert}

%
%
%
%
%
%
%
\def\mactrue{t}
\def\figI{I}
\newcount\figscale
\newdimen\figwidth
\newdimen\figheight
\ifx\figflag\figI
\def\ifigure#1#2#3#4#5#6{
\figscale=#6
\figwidth=#4truein
\figheight=#5truein
\ifx\answ\bigans
    \multiply\figscale by 2
\else
    \multiply\figscale by 12 \divide\figscale by 5
\fi
\midinsert
\centerline{\vbox to \the\figheight{
\hrule width \the\figwidth height 0pt depth 0pt \vfill
\ifx\macflag\mactrue
    \special{illustration #3.eps scaled \the\figscale}
\else
    \multiply\figscale by 2 \divide\figscale by 5
    \includegraphics{#3.ps}
\fi
}}
\narrower\narrower\noindent #1: #2
\endinsert
}
%
\ifx\macflag\mactrue\else

\fi
%
\else
\def\ifigure#1#2#3#4#5#6{\ifig{#1}{#2}{#5}}
\fi
%
\def\pa{\partial}
\def\pd#1#2{{\partial #1\over\partial #2}} 
\def\dbd#1{{\del \over \pa{#1}}}
\def\der#1#2{{{\pa{#1}} \over {\pa{#2}}}}
\def\rr{{\tilde r}}
\def\rrho{{\tilde \rho}}
\def\rrb{\overline{\rr}}
\def\rb{\overline{\vphantom\i {r}}}
\def\tb{\overline{\vphantom\i {t}}}
\def\aa{{\tilde a}}
\def\mh#1#2{\widehat {\cal M}_{#1,#2}}
\def\ph#1#2{\widehat {\cal P}_{#1,#2}}
\def\mgs#1#2{{\cal M}_{#1,#2}}
\def\pgs#1#2{ {\cal P}_{#1,#2}}
\def\osp#1#2{$Osp(#1,#2)$}
\def\s{\sigma}
\def\th{\theta}
\def\b#1#2{{b}^{(#2)}_{#1}}
\def\bb#1#2{{\bar b}^{(#2)}_{#1}}
\def\we{\wedge}
\def\rrmeas{[\dd \rr\we\dd\rrb|\dd\rrho]}
\def\tmeas{[\dd t\we\dd\tb|\dd\tau]}
\def\rtmeas{\bigl[\dd r\we\dd\rb\we\dd t\we\dd\tb|\dd\rho\dd\tau\bigr]}
\def\db#1{\delta[\beta_{#1}]}

\def\dbt#1#2{\delta[\beta^{(#2)}_{#1}]}
\def\dbpt#1#2{\delta'[\beta^{(#2)}_{#1}]}
\def\B#1{B[\s_*({#1})]}

\def\bet#1#2{\beta^{(#2)}_{#1}}
\def\mha{-{1\over2}}
\def\kpkq#1#2{\ket{#1}^P\otimes\ket{#2}^Q}
\def\kpkqm#1#2{\ket{#1}^P_M\otimes\ket{#2}^Q_M}

\def\pp{\tilde p}
\def\z#1{\zeta_{#1}(\cdot)}
\def\zc#1{\check{\zeta}_{#1}(\cdot)}
\def\cd{(\cdot)}
\def\G#1#2{G^{(#1)}_{#2}}
\def\L#1#2{L^{(#1)}_{#2}}
\def\Lb#1#2{\bar{L}^{(#1)}_{#2}}
\font\llcaps=cmcsc10
\def\BER{\hbox{\llcaps ber$\,$}}
\def\det{\hbox{det$\,$}}                
\def\sdet{\hbox{sdet}\,}                
%
%
%
\lref\Kutconf{D. Kutasov, ``Geometry on the space of conformal field
theories and contact terms,'' Phys. Lett. {\bf B220} (1989) 153.}%
\lref\GrSei{M. Green and N. Seiberg, ``Contact interactions in
superstring theory," Nucl. Phys. {\bf B299} (1988) 559.}%
\lref\DWTG{R. Dijkgraaf and E. Witten, ``Mean field theory, topological
field theory, and multi-matrix models,'' Nucl. Phys. {\bf B342} (1990)
486.}%
\lref\DESRST{J. Distler and P. Nelson, ``The dilaton equation in
semirigid string theory," Nucl. Phys. {\bf B366} (1991) 255.}%
\lref\EWTG{E. Witten, ``On
the structure of the topological phase of two-dimensional gravity,''
Nucl. Phys. {\bf B340} (1990) 281.}%
\lref\VVTG{E. Verlinde and H. Verlinde, ``A solution of two-dimensional
topological gravity,'' Nucl. Phys. {\bf B348} (1991) 457.}%
\lref\semirig{J. Distler and P. Nelson, ``Semirigid supergravity,''
Phys. Rev. Lett. {\bf66} (1991) 1955.}%
\lref\CIGVO{P. Nelson, ``Covariant insertions of general vertex operators,"
Phys. Rev. Lett. {\bf62} (1989) 993.}%
\lref\EFEFS{H.S. La and P. Nelson, ``Effective field equations for
fermionic strings," Nucl. Phys. {\bf B332} (1990) 83.}%
\lref\TCCT{J. Distler and P. Nelson, ``Topological couplings and
contact terms in 2d field theory,'' Commun. Math. Phys. {\bf138} (1991) 273.}%
\lref\JDTG{J. Distler, ``2-d quantum gravity, topological field theory,
and the multicritical matrix models,'' Nucl. Phys. {\bf B342} (1990) 523.}%
\lref\AGNSV{L. Alvarez-Gaum\'e, C. Gomez, P. Nelson, G. Sierra, and C.
Vafa, ``Fermionic strings in the operator formalism," Nucl. Phys.
{\bf B311} (1988) 333.}%
\lref\AGETAL{L. Alvarez-Gaum\'e, C. Gomez, G. Moore, and C.
Vafa, ``Strings in the operator formalism," Nucl. Phys.
{\bf B303} (1988) 455.}%
\lref\ARS{J. Atick, J. Rabin and A. Sen, Nucl. Phys. {\bf B299} (1988)
279\semi
H. Verlinde, ``A note on the integral over the fermionic supermoduli,"
Utrecht preprint, THU-87/26 (1987).}%
\lref\AMS{J. Atick, G. Moore and A. Sen, ``Catoptric tadpoles," Nucl. Phys.
{\bf
B307} (1988) 221.}
\lref\Meas{P. Nelson, ``Lectures on supermanifolds \& strings,"
Proceedings of the TASI Summer School, (1988).}%
\lref\JPfact{J. Polchinski, ``Factorization of bosonic string amplitudes,''
Nucl. Phys. {\bf B307} (1988) 61.}%
\lref\JPvert{J. Polchinski, ``Vertex
operators in the Polyakov path integral,'' Nucl. Phys. {\bf B289} (1987) 465.}%
\lref\LSMS{P. Nelson, \cmp{``Lectures on strings and moduli space,''} Phys.
Reports {\bf149} (1987) 304.}%
\lref\EUGENE{E. Wong, ``Recursion relations in semirigid topological
gravity'', Penn U. preprint UPR-0491T, Nov. 1991.}%
\lref\DM{P. Deligne and D. Mumford, ``The irreducibility of the space of curves
of given genus," IHES Publ. Math. {\bf 36} (1969), 75.}%
\lref\KM{F. Knudsen and D. Mumford, ``The projectivity of the moduli
space of stable curves,~I," Math. Scand. {\bf 39} (1976) 19.}%
\lref\MYTOME{M. Doyle, ``Dilaton contact terms in the bosonic and
heterotic strings," Princeton preprint PUPT-1296, hepth 9201076, submitted
to Nucl. Phys. {\bf B}.}%
\lref\SZ{H. Sonoda and B. Zweibach, ``Covariant closed string theory cannot be
cubic," Nucl. Phys. {\bf B336} (1990) 185.}%
\lref\COHN{J. Cohn, ``Modular geometry of superconformal field theory,"
Nucl. Phys. {\bf B306} (1988) 239.}%
\lref\DELIGNE{P. Deligne, unpublished (1987).}%
\Title{PUPT--1312}
{World-Sheet Supersymmetry Without Contact Terms$^\star$}

\centerline{Jacques Distler and Mark D. Doyle}\smallskip
\centerline{Joseph Henry Laboratories}
\centerline{Princeton University}
\centerline{Princeton, NJ \ 08544 \ USA}
\bigskip\bigskip

\footnote{}{{\parindent=-10pt\par $\star$
\vtop{
\hbox{Email: {\tt distler@puhep1.princeton.edu, mdd@puhep1.princeton.edu} .}
\hbox{Research supported by NSF grant PHY90-21984.}
     }     }}

Green and Seiberg showed that, in simple treatments of fermionic string theory,
it is necessary to introduce contact interactions when vertex operators
collide.  Otherwise,
certain superconformal Ward identities would be violated.
In this note, we show how these contact terms arise naturally when proper
account is taken of the superconformal geometry  involved when
punctures collide. More precisely, we show that there is no contact term at
all!
Rather, corrections arise to the ``na\"\i ve" formula when the boundary of
moduli space is described correctly.

\Date{3/92}                 
\def\test{T}\ifx\prlmode\test \baselineskip22pt\fi

The existence of contact terms in string theory seems, after a moment's
thought,
to be somewhat paradoxical. In the {\it stable compactification}
of moduli space
\refs{\DM,\KM},
punctures ({\it i.e.}~the locations at which vertex operators are inserted)
are, {\it by definition} not allowed to collide. Rather, as two punctures
approach each other, we can by a conformal  transformation take the punctures
to
remain separated while  a piece of the surface surrounding them pinches off
(\tfig\fcoll). If the punctures never collide, there would seem to be little
role for a $\delta$-function contact term. A more sophisticated person might
replace the notion of a contact term when punctures coincide with that of a
contribution
to the string measure which is ``{\sl $\delta$-function-concentrated
on the boundary of moduli space corresponding to the pinched surface}." But
this formulation seems rather strained and hard to derive from conformal field
theory. After all, conformal field theory amplitudes are  supposed to be
real-analytic
functions of the moduli, which a $\delta$-function certainly is not.

\ifigure\fcoll{Colliding punctures.}{colli}{4}{1.25}{100}

Fortunately, it now appears that
``contact interactions" have a perfectly natural
home in string theory. What appear to be contact interactions arise
as a consequence of certain global features of the moduli space of punctured
Riemann surfaces. The important point is that in order to define the operators
which go into a conformal field theory correlation function, we need,
implicitly, a normal-ordering prescription or, what amounts to the same thing,
a choice of a local
coordinate at the insertion point. Actually, we need a family
of such local coordinates which vary as we vary the moduli of the surface. Our
standard beliefs about the properties of string amplitudes obtain when the
coordinate families vary holomorphically with the moduli. Unfortunately, there
are obstructions  to finding such a holomorphic coordinate family globally on
moduli space.  We are forced to allow our normal-ordering prescription to vary
non-holomorphically with the moduli.

One consequence of this phenomenon is that the one-point function of the
zero momentum dilaton vertex
operator (which na\"\i vely decouples) is not zero, but rather is
proportional to
the Euler characteristic of the Riemann surface
\refs{\JPfact,\CIGVO,\EFEFS}.
In the presence of other vertex operator insertions, one needs a more
refined treatment of the behavior of coordinate families as punctures
collide. This leads to the dilaton insertion being proportional to the Euler
characteristic of the {\it punctured} Riemann surface
\TCCT. In other words, the dilaton insertion counts the powers
of the string coupling in the correlation function, or, if you
wish, there is a low-energy theorem $D\sim\lambda\pd{}{\lambda}$. For the
bosonic string, this is true only for the first variation with respect to
$\lambda$ (one dilaton insertion), but for the heterotic string, it holds to
all
orders (multiple dilaton insertions) \MYTOME.
In similar fashion, one can recover the
``dilaton" \DESRST\ and ``puncture"
\EUGENE\ equations of topological gravity, postulated
in \refs{\DWTG,\VVTG}. The key point
to be emphasized here is that none of these results involve mysterious
$\delta$-functions at the boundary of moduli space. The string measure is
perfectly smooth and, aside from physical on-shell poles, vanishes as one
approaches the boundary of moduli space.

Another sort of contact term which one encounters in fermionic string theory
is that proposed by Green and Seiberg \GrSei.
The ``integrated" form of a NS vertex operator ({\it i.e.}~when one thinks
of the vertex operator as a 2-form to be integrated over the Riemann surface)
is naturally in the $(0)$-picture. However, if one na\"\i vely takes the
OPE of two
such vertex operators as the approach each other, the result violates the
superconformal Ward identities.  The ``solution" is to add a $\delta$-function
contact term which restores the Ward identities. Green and Seiberg showed that
(at least for free field theories) this contact term could be explained by
including the auxiliary fields, $F$, necessary to close the world-sheet
supersymmetry
algebra off-shell. The equation of motion for the auxiliary field is simply
$F=0$, and the auxiliary field doesn't propagate -- its two-point function is
just a $\delta$-function. But, noted Green and Seiberg, if we include the
contribution of the auxiliary field to the $(0)$-picture vertex operator, the
extra terms which arise from contracting the auxiliary fields when the
vertex operators collide precisely reproduce the needed contact terms.

This mechanism, though pretty, has some drawbacks. First, it only applies to
field theories in which one knows an off-shell auxiliary field structure. This
leaves out a large class of interesting theories which are only superconformal
on-shell. One would like to be able to treat more general superconformal
theories. Second, by explicitly invoking off-shell features of the theory, one
is {\it potentially} opening a Pandora's box of violations of superconformal
invariance. For free theories, it appears that
one can easily snap shut the lid after extracting
 the desired contact terms. For more general theories, even if their auxiliary
field structure is known, one may not be so lucky.

In the operator formalism, a vertex operator is naturally thought of as a
0-form inserted at a puncture; it always appears in the $(-1)$-picture.
The picture-changing operator which moves it into the $(0)$-picture appears
when one integrates over the supermodulus associated to the location of the
puncture. When two vertex operators collide, we should not be too cavalier in
our treatment of picture-changing. We shall see in this paper that in a
more careful treatment of the collision of two punctures, extra terms arise in
addition to the usual ``picture-changing" terms. These are precisely what is
necessary to restore the superconformal Ward identities. Thus in a careful
treatment of the degenerating surface (\fcoll), no extra ``contact terms" are
necessary; all of the necessary corrections are automatically taken into
account. All of this can be carried through for an arbitrary superconformal
field theory. Indeed, we won't specify precisely what superconformal theory we
are working with.

\newsec{Preliminaries}
\subsec{The Measure}
As discussed above, the definition of the normal-ordered
vertex operators, and hence of the
string measure, depends on a choice of local coordinate at the insertion
points.
The operator formalism provides a nice framework for keeping track of this
dependence and constructing explicit expressions for the superstring measure.
The extra data of a local coordinate at each puncture fits
together into an infinite dimensional fiber bundle $\ph gn\to\mh gn$. Choosing
 a family of local coordinates at the punctures amounts to choosing a section
$\sigma: \mh gn\to\ph gn$. The measure on $\mh gn$ is defined by
\eqn\hmeas{\eqalign{\Omega(v_1,v_2,\ldots,\bar v_{3g-3+n},
     &\nu_1,\nu_2,\ldots,\nu_{2g-2+n})=\cr
 \bra{\Sigma,(z_1,\th_1),&\ldots}B[\sigma_* v_1]
\ldots B[\sigma_* \bar v_{3g-3+n}]\cr&\times\delta(B[\sigma_*\nu_1])\ldots
\delta(B[\sigma_*\nu_{2g-2+n}])
         \ket{\psi_1}_{P_1}\otimes \cdots\otimes\ket{\psi_n}_{P_n}.}}
where $\sigma_*v_i$ is the push-forward of a tangent vector to $\mh gn$ to a
tangent
vector to $\ph gn$. This construction, and the above notation, are described in
great detail in \refs{\AGNSV,\EFEFS} and \MYTOME.

The tangent space to $\ph gn$ has an elegant presentation in terms of
Schiffer variations. It has the structure of a direct sum of  $n$
copies of the Neveu-Schwarz algebra (one for each puncture) modulo a
certain ``Borel" subalgebra. The generators of this algebra are given in
terms of the local coordinates at the puncture by
 \eqn\nsa{\eqalign{
   l_n& \leftrightarrow
-z^{n+1}\dbd{z}-\half(n+1)z^n\theta\dbd{\theta}\cr
     g_k& \leftrightarrow \half z^{k+\ha}
          \bigl(\dbd{\theta} -\theta\dbd{z}\bigr).}}
An infinitesimal change in the coordinate used at a puncture is given by the
action of the ``vertical" tangent vectors $l_n,\ g_{n+\ha},\ n\geq0$. These
induce a change of the state associated to the punctured surface given
by
\eqn\NSA{\bra{\Sigma, z-\epsilon z^{n+1}-\half\hat{\epsilon}\theta
z^{k+\ha}}=\bra{\Sigma, z}\bigl(1+\epsilon L_n +\hat{\epsilon}
G_k\bigr)+\cdots .}
Again, details may be found in  \refs{\AGNSV,\EFEFS,\MYTOME}.

\subsec{The Coordinate Family}
The geometry appropriate for computing contact terms in the operator
formalism is now well-established. The heterotic case was treated in
detail in \MYTOME\ and we will borrow the results of some of the
calculations presented there. Here we will be content to give a brief
account of the geometry and the reader should consult \MYTOME\ for a more
complete account.

When the points $P$ and $Q$ are widely separated, it is appropriate to
use coordinates at one point that are independent of the location of the
other. A convenient choice of coordinate is the so-called superconformal
normal-ordered (SCNO) coordinate \EFEFS, and letting $(\rr,\rrho)$ be the
location of $P$, we take as coordinates
\eqn\SCNO{\zeta_P(\cdot)= z(\cdot)-\rr-\rrho\th(\cdot),\qquad
                  \check{\zeta}_P(\cdot)=\th(\cdot)-\rrho\quad.}
The point $P$ should be thought of as the place where $\zeta_P(\cdot)$ and
$\check{\zeta}_P(\cdot)$ vanish. $\z P$ and $\zc P$ are coordinates that
are superconformally related to $z\cd$ and $\th\cd$. For a SCNO
coordinate, $\zc P$ is just $D\z P$, where $D$ is the super-derivative
$\dbd{\th}+\th\dbd{z}$. Coordinates at $Q$, located at, say, $(r,\rho)$,
can be given by similar expressions, with $\rr$ and $\rrho$ replaced by
$r$ and $\rho$.

In conformal field theories, the collision of two punctures is replaced
by the conformally equivalent picture of the two punctures located on a
sphere pinching off from the rest of the surface. In this region SCNO
coordinates are not appropriate. Instead, the coordinates of the point
$P$ will depend on the location of $Q$, and they are given by a plumbing
fixture construction \COHN. This construction is obtained by beginning with a
three-punctured super-sphere. We can use \osp{2}{1} symmetry to fix
the three bosonic coordinates of the punctures at 0, 1, and $\infty$.
Only two of the three fermionic coordinates can be fixed though; the
remaining one is the one odd modulus of the three-punctured
super-sphere, which we denote $\tau$. Thus we place our three punctures
at (0,0), (1,$\tau$) and ($\infty$,0). The point at $\infty$ is sewn onto
the rest of the surface at the old location of $Q$.
If $w$ and $\xi$ are the
coordinates on the super-sphere, we can write the most general
coordinates that vanish at $P$ and $Q$ as
\eqn\zercor{w-\aa_1 \tau w \xi + \aa_2 w^2 -\aa_3 \tau w^2 \xi +\cdots}
at (0,0) and
\eqn\onecor{(w-1+ \tau \xi)-  a_1  \tau (w-1+ \tau \xi) (\xi - \tau)
+a_2 (w-1+ \tau \xi)^2 -  a_3 \tau (w-1+ \tau \xi)^2 (\xi -
\tau)+\cdots } at (1,$\tau$). We have just given the even coordinates
here. The odd coordinates are found by demanding that the coordinate
transformation from $(w,\xi)$ be superconformal \MYTOME. The minus signs
are for later convenience. At $\infty$ we can simply use $-1/w$ and
$\xi/w$. These coordinates are sewn to the coordinates $\z{Q}$ and
$\zc{Q}$ by the identifications
\eqn\sewing{w={\z{Q}\over t^2}\quad,
             \quad\quad \xi=-{\zc{Q}\over t}\qquad.}
Thus, we have as coordinates for $P$ and $Q$, when they are close
together,
\eqn\sewncor{\eqalign{\phi_Q(\cdot)=&{\zeta_Q(\cdot)\over t^2}+
 \aa_1\tau {\zeta_Q(\cdot) \check{\zeta_Q}(\cdot) \over t^3}+
                         \aa_2 {\zeta_Q(\cdot)^2 \over t^4} +\cdots\cr
\phi_P(\cdot)=&\Sigma(\cdot) +a_1 \tau\Sigma(\cdot) \check\Sigma(\cdot)
                  +a_2 \Sigma(\cdot)^2 +\cdots\,,}}
with
\eqn\Sigdef{
    \Sigma(\cdot)=\bigl({\zeta_Q(\cdot) \over t^2} -1 -
                {\tau \check{\zeta_Q}(\cdot) \over t}\bigr),\qquad
\check\Sigma(\cdot)=\bigl({\check{\zeta_Q}(\cdot)\over t}+\tau\bigr)\quad.}

Notice that the coordinates for $P$ are now given in terms of $r$,
$\rho$, $t$, and $\tau$ instead of $\rr$ and $\rrho$.
These are the coordinates for supermoduli space
appropriate to describe the neighborhood of
this boundary of
supermoduli space.
On the overlap between the two patches of supermoduli space, the change of
coordinates from $r,\rho,t,\tau$ to $r,\rho,\rr,\rrho$
is found by demanding that $\phi_P(P)$ (and its superconformal partner
$\check{\phi}_P(P)$) vanish at the same point where the coordinate family in
the other patch are centered. This happens when $\zeta_Q(P)=t^2$ and
$\check{\zeta}_Q(P)=-t \tau$. We find
\eqn\rteq{\rr= r+t^2 +t \rho\tau,\qquad \rrho=\rho-t\tau\quad.}
The non-split nature of this transformation is ultimately responsible
for the corrections to the ``na\"\i ve'' calculation that restores the
Ward identities$^\#$\footnote{}{{\parindent= -10 pt\par
$^\#$ The role of non-split coordinate transformations
on supermoduli space in this regard
was also noted by H. Verlinde
\ref\rHVpc{H. Verlinde, private communication.}.}}.
 This will be seen explicitly below.

Finally, to get a coordinate family appropriate for both when the
punctures are widely separated and when they are close together,
one should interpolate smoothly between the two behaviors outlined above.
We do this by introducing a smooth function  $f(|t|)$ that goes from
0 to 1 as $|t|$ goes from 0 to $\infty$, and using a linear
interpolation. Thus, the coordinates that we will use are
\eqn\slice{\eqalign{\sigma_P(\cdot)=&{f(|t|) \over t^2}
                 \zeta_P(\cdot)+(1-f(|t|)) \phi_P(\cdot)\cr
                 \sigma_Q(\cdot)=&{f(|t|) \over t^2}\zeta_Q(\cdot) +
                    (1-f(|t|)) \phi_Q(\cdot)\,.}}
The precise shape of the function $f$ will drop out of the calculation. Any
coordinate family which interpolates between the required behavior at large
$|t|$ and at small $|t|$ will yield the same value for the measure.
(In the case of dilaton correlation functions, the measure itself does depend
on the interpolating function $f$; only its integral is independent of $f$.)

\newsec{The calculation}

The state $\bra{\Sigma,\s_P,\s_Q}$ at $r=\rb=0$ can be expanded as
\eqn\state{\eqalign{
\bra{\Sigma,\s_P,\s_Q}=&\bra{\Sigma,\s_{P_0},\s_{Q_0}}
\bigl(1+\rho\tau t \L{P}{-1}+\rho\tau\bigl({2 p_2-p_1\over t}\bigr)
\L{P}{0} +2\bigl(\rho - \tau t\bigr)\G{P}{\mha}+\cdots\bigr)\cr
 & \times\bigl(1-{\rho \tau \tilde{p}_1 \over t}\L{Q}{0} + 2\rho
\G{Q}{\mha}+\cdots\bigr)
       t^{2 (\L{P}{0}+\L{Q}{0})} \tb^{2(\Lb{P}{0}+\Lb{Q}{0})}
}}
where $p_1=a_1 (1-f)$, $\pp_1=\aa_1 (1-f)$, and $p_2=a_2 (1-f)$.
$\s_{P_0}$ and $\s_{Q_0}$ are the equal to $\s_P$ and $\s_Q$ with $\rho$
and $\tau$ set to zero, and rescaled by a factor of $t^2$.
The ghost insertions that result from computing the
pushforwards of the tangent vectors associated to the moduli
corresponding to the locations of
the points $P$ and $Q$ by the coordinate family are given in \MYTOME.
The measure is
formed by multiplying these six contributions together (including
delta functions for the insertions for $\rho$ and $\tau$).
The insertions are (evaluated at
$r=\rb=0$)
\eqn\hetghost{\eqalign{
\B{\dbd{r}}=&
   -\bigl({1 \over t^2}\bigr)\b{-1}{P}+
   \biggl({2 p_1 \tau \over t^2}\biggr)\bet{\mha}{P}-
  \bigl({1\over t^2}\bigr)\b{-1}{Q}+
   \biggl({2 \pp_1 \tau \over t^2}\biggr)\bet{\mha}{Q}+\cdots
\cr
\B{\dbd{\rb}}=&
  -\bigl({1\over\tb^2} \bigr)\bb{-1}{P}-
   \bigl({1\over\tb^2}\bigr)\bb{-1}{Q}+\cdots
\cr
\B{\dbd{\rho}}=&
   \biggl(-{\rho\over t^2}+{2 \tau \over t}\biggr)\b{-1}{P}+
   \biggl({2\over t}+{2 p_1 \rho \tau\over t^2}\biggr)\bet{\mha}{P}-\cr
  &\biggl({\rho \over t^2}\biggr)\b{-1}{Q}+
   \biggl({2\over t}+{2 \pp_1 \rho \tau \over t^2}
          \biggr)\bet{\mha}{Q}+\cdots\cr
\B{\dbd{t}}=&
   -\biggl({2\over t}\biggr)\b{-1}{P}+
   \biggl({2 \tau\over t}+{4 p_1 \tau\over t}\biggr)\bet{\mha}{P}+\cdots
\cr
\B{\dbd{\tb}}=&
   -\biggl({2\over \tb}\biggr)\bb{-1}{P}+\cdots
\cr
\B{\dbd{\tau}}=&
   -\bigl(\tau\bigr)\b{-1}{P}-
  \bigl(2\bigr)\bet{\mha}{P}+\cdots
\cr
}}
where the dots represent higher terms ($ b_n, n\ge0$ and
$ \beta_n, n\ge \ha$)
that annihilate strong physical states (SPSs).

To insert a SPS at $P$ and at $Q$, there are two possible contributions
to the measure from the above insertions:
\eqn\bcont{
\b{-1}{P} \bb{-1}{P}\dbt{\mha}{P}\b{-1}{Q} \bb{-1}{Q}\dbt{\mha}{Q}\ ,\qquad
\b{-1}{P} \bb{-1}{P}\dbt{\mha}{P}\b{-1}{Q} \bb{-1}{Q}
     \bet{\mha}{Q}\dbpt{\mha}{Q}\ .}
Using the formal identification of $\beta \delta'(\beta)$ and
$-\delta(\beta)$, we easily find the contribution from the
insertions in \hetghost\ to be
\eqn\bmeas{{t\over |t|^6} \b{-1}{P} \bb{-1}{P}\dbt{\mha}{P}\b{-1}{Q}
\bb{-1}{Q}\dbt{\mha}{Q},} where there has been some cancellation between
the two different contributions in \bcont.

Thus we have
\eqn\contact{\eqalign{\int &\rtmeas
 \ \bra{\Sigma,\sigma_P,\sigma_Q}\cdots
\B{\dbd{r}}\B{\dbd{\rb}}\delta\big[\B{\dbd{\rho}}\big]\cr
     &\qquad\times \B{\dbd{t}}\B{\dbd{\tb}}\delta\big[\B{\dbd{\tau}}\big]
\kpkq{\psi}{\psi}=\cr
\int &\rtmeas\ \bra{\Sigma,\s_{P_0},\s_{Q_0}}\cr
&\qquad\times\bigl(1+\rho\tau t\L{P}{-1}+
  \rho\tau\bigl({2 p_2-p_1\over t}\bigr)\L{P}{0}
 +2\bigl(\rho - \tau t\bigr)\G{P}{\mha}\bigr)\cr
 & \qquad\times\bigl(1-{\rho \tau \tilde{p}_1 \over t}\L{Q}{0} + 2\rho
\G{Q}{\mha}\bigr)
       t^{2 (\L{P}{0}+\L{Q}{0})} \tb^{2(\Lb{P}{0}+\Lb{Q}{0})}\cr
&\qquad\times\bigl({t\over
|t|^6}\bigr)\b{-1}{P}\bb{-1}{P}\dbt{\mha}{P}\b{-1}{Q}
\bb{-1}{Q}\dbt{\mha}{Q}\kpkq{\psi}{\psi}.
}}
The state $b_{-1}\bar{b}_{-1}\db{\mha}\ket{\psi}$ has $L_0=\ha$ and
$\bar{L}_0=1$ and thus, picking off the term proportional $\rho\tau$,
\contact\ becomes
\eqn\contacti{\eqalign{\int&\rtmeas\cr
\bra{\Sigma,\s_{P_0},\s_{Q_0}}& \rho\tau\biggl(t \L{P}{-1}+
    { (2 a_2 -a_1-\tilde{a}_1)(1-f) \over 2 t}
+4t\G{Q}{\mha}\G{P}{\mha}\biggr)\cr
&\times\bigl({|t|^2\over t}\bigr)
 \b{-1}{P}\bb{-1}{P}\dbt{\mha}{P}\b{-1}{Q}
\bb{-1}{Q}\dbt{\mha}{Q}\kpkq{\psi}{\psi}.
}}
At first sight, the term proportional to $(1-f)$ looks like bad news, since it
depends on the shape of the function $f$, and hence on the details of the
coordinate family chosen. Actually, the coefficients
$a_i$ and $\aa_i$ are not to be taken to be arbitrary. As discussed in \MYTOME,
a unique choice for the coefficients is singled out by demanding that dilaton
insertions behave correctly. There one found that $a_1=a_2=\aa_1=-\aa_2=-1/4$.
(Actually, the $\aa_i$ were not determined in \MYTOME, but a calculation
exactly analogous to the ones presented there fixes these coefficients as
well.)
With these values for the coefficients, the offending term vanishes, and all
dependence on the shape of the coordinate family chosen drops out.
{\it Any} coordinate family which interpolates to the asymptotic behavior
\zercor,\onecor\ gives the same result for the measure
\eqn\contactii{\eqalign{\int\rtmeas&\ \bra{\Sigma,\s_{P_0},\s_{Q_0}}
\rho\tau\bigl(\L{P}{-1} +4\G{Q}{\mha}\G{P}{\mha}\bigr)\cr
&\times |t|^2
 \b{-1}{P}\bb{-1}{P}\dbt{\mha}{P}\b{-1}{Q}
\bb{-1}{Q}\dbt{\mha}{Q}\kpkq{\psi}{\psi}.
}}

Our demand that the coordinate family exhibit a certain asymptotic behavior
as we approach the boundary of moduli space has both a physical motivation
(it gives the correct behavior for dilaton correlation functions) and a
mathematical motivation. When we think about the plumbing construction
as gluing a certain {\it fixture} -- a 3-punctured sphere {\it with}
coordinates -- onto the rest of the surface, it makes sense to demand that the
resulting sewn surface be the same, regardless of which puncture we chose to
attach to the rest of the surface. In other words, we demand invariance under
(a subgroup of) m\oe bius transformations of the 3-punctures sphere. As in
the bosonic string \refs{\TCCT,\SZ}, this leads to a definite choice for the
first few coefficients in \zercor,\onecor.
Still, one might be puzzled that the measure formed with SPSs should
be sensitive at all to the asymptotic behavior of the coordinate family.
We will defer a discussion of this point to section 3.

Of the two terms in \contactii, the $\G{Q}{\mha}\G{P}{\mha}$ is precisely what
is expected from a na\"\i ve analysis of the collision of two vertex operators.
Indeed, the $\G{P}{\mha}$ combines with the $\dbt{\mha}{P}$ to form a
picture-changing operator, which converts the state at $P$ from the $(-1)$
 to the $(0)$ picture (and similarly for the state at $Q$). The $\L{P}{-1}$
term
is new and, as we shall see, plays the role of the ``contact term"
postulated by Green and Seiberg.

First, let us see that it makes sense by examining the transformation of
the measure under the coordinate transformation in \rteq.
There are three contributions to the change. The first is the how the
integration measure $\rrmeas$ transforms to $\tmeas$. The change of
variables requires the introduction of a Jacobian (Berezinian),
\eqn\berez{\rrmeas=\sdet\biggl(\matrix{\der{\rr}{t}&\der{\rrho}{t}\cr
                         \der{\rr}{\tau}&\der{\rrho}{\tau}}\biggr)
  \det \biggl(\der{\rrb}{\tb}\biggr)\tmeas.}
Using \rteq, this results in
\eqn\berezi{\rrmeas={4 |t|^2\over t} \bigl(1+{\rho \tau \over 2 t}\bigr)
                     \tmeas.}
But the $B$ insertions \hetghost\ also change, introducing a compensating
Jacobian in  \bmeas. Finally,
changing variables to $t$ and $\tau$ also changes the state
$\bra{\Sigma}$. Equation \NSA\ gives
\eqn\stch{\eqalign{
  \bra{\Sigma, z-\rr+\rrho\th}=&
            \bra{\Sigma,z-\rr}\bigl(1- 2\rrho G_{\mha}\bigr)
              = \bra{\Sigma,z-(r+t^2)-t \rho\tau-(\rho-t\tau)\th}\cr
  =&\bra{\Sigma,z-(r+t^2)}\bigl(1+t\rho\tau L_{-1}
                             +2(\rho-t\tau) G_{\mha}\bigr).}}
In this formula, we see the consequences of the non-split nature of \rteq.
The $L_{-1}$ term would not be there for some split transformation.
So, though the measure is perfectly continuous, the appearance of the extra
term in \contactii\ is a consequence of expanding it in coordinates appropriate
to the neighborhood of the boundary of moduli space when the two punctures
collide.

Recall that our goal is to perform an operator product expansion, which
in this context amounts to rewriting the state that arises from the
collision of two operators as the insertion of a single operator at the
collision point. The result in Green and Seiberg was that the na\" \i ve OPE
resulted in a state that was not the highest component of a superfield and
hence violated supersymmetry. Restoring supersymmetry required adding a contact
term. However, as we
will shortly see, the state that we find {\it is} the highest component of a
superfield inserted at the collision point, and thus, no contact term is
needed.
It is precisely the
``extra" $L_{-1}$ term that restores supersymmetry. The key point is to
appreciate what ``taking the OPE" means in the operator formalism: take the two
states at $P$ and $Q$, contract them with the state associated to the
3-punctured sphere, and insert the state that results  at the neck where the
3-punctured sphere is joined onto the rest of the surface (\fcoll). In doing
so,
it is vital to use the coordinates appropriate to this superconformal geometry.
In particular, to see that the state inserted at the neck is the highest
component of a superfield, we need to use a contour-deformation, or more
pedantically, to make use of the Borel-ambiguity associated to the 3-punctured
sphere.
{}From the Neveu-Schwartz algebra, we have
$$L_{-1}=2\{G_{\mha},G_{\mha}\}=4 G^2_{\mha}.$$
Thus, \contactii\ becomes
\eqn\contactiii{\eqalign{\int\rtmeas&\   4\rho\tau |t|^2
\bra{\Sigma,\s_{P_0},\s_{Q_0}}
\bigl(\G{P}{\mha}+\G{Q}{\mha}\bigr)\G{P}{\mha}\cr
&\times\b{-1}{P}\bb{-1}{P}\dbt{\mha}{P}\b{-1}{Q}
\bb{-1}{Q}\dbt{\mha}{Q}\kpkq{\psi}{\psi}.}}
The factor of
$(\G{P}{\mha}+\G{Q}{\mha})$ is equivalent (up to the Borel-ambiguity
associated to the 3-punctured sphere) to an insertion $\G{N}{\mha}$,  on the
neck of the sphere that is pinching off. Effectively, we can replace the
sum of two contours surrounding $P$ and $Q$ with one contour surrounding both
and
deformed this contour onto the neck (\tfig{\cont}).

\ifigure{\cont}{Contour manipulations.}{contours}{5}{1.25}{100}

The $b$ and
$\delta(\beta)$ insertions have removed the ghost  dependence in our SPSs and
we
are left with a state  $\ket{\tilde{\psi}}_M$ in the matter theory, inserted on
the neck, given by
\eqn\psit{\ket{\tilde{\psi}}_M=\G{N}{\mha}\ket{\sigma_N}\bra{\sigma_P,\sigma_Q}
\G{P}{\mha}\kpkqm{\psi}{\psi}\quad.}
Here $\ket{\sigma_N}\bra{\sigma_P,\sigma_Q}$ is the state associated to the
3-punctured sphere, viewed as an element of $\CH\otimes\CH^*\otimes\CH^*$.
Since
$\ket{\tilde{\psi}}_M$ is the highest component of a  superfield ({\it i.e.}~it
is $G_{\mha}$ acting on the lowest component of a superfield), it is
supersymmetric.

Thus we have maintained world-sheet supersymmetry {\it without} adding an
explicit contact term when the two vertex operators coincide. The correction to
the ``na\" \i ve" result (of taking the OPE of two (0)-picture vertex
operators)
arises not from an explicit contact term, but rather from a proper account of
the superconformal geometry associated to the collision of two punctures.
Since the correction is proportional to $\L{P}{-1}$, it looks like a total
derivative in the modulus associated to the location of $P$. Hence its effects
can be {\it simulated} by adding a $\delta$-function contact term
at the boundary of moduli space.

\newsec{Concluding Remarks}

One slightly puzzling feature of our calculation is that the dependence of
\contacti\ on the coordinate family used to construct the measure dropped out
only for coordinate families with the correct asymptotic behavior. In the
bosonic
string, one can prove that the measure for SPSs  is completely insensitive to
the
coordinate family chosen. But the argument rests on the fact that  $\Omega$ is
the pullback of a differential form on $\pgs gn$ which, for SPSs,
 has the
property that a) it is constant along the fibers of $\pgs gn\to \mgs gn$ and b)
it annihilates the vertical directions. In the super case, however, $\Omega$
is in no sense a pullback of a differential form on $\ph gn$. Rather
\ref\rRoth{M. Rothstein, Trans. AMS. {\bf 299} (1987) 387.}, it is a
section of the sheaf $\BER$ of Berezin forms on $\mh gn$ and has no
invariantly-defined precursor ``upstairs" on $\ph gn$. So the corresponding
argument cannot be made in the super case. Nevertheless, we have seen that
merely demanding the correct asymptotic behavior of the coordinate family is
enough in this case to remove all dependence on the ``shape" of the coordinate
family chosen. Perhaps a generalization of the arguments of \MYTOME\ would
allow this to be extended globally over moduli space.
If not, then it might still be the case that these unwanted
contributions are actually total derivatives in the moduli and, hence,
decouple anyway, although that is hard to see explicitly in this
particular calculation. Finally, let us note that
the possible dependence of the string measure on the coordinate family is quite
distinct from the ``integration ambiguity" raised in the literature (see
\refs{\ARS,\AMS} and references therein) a few years ago. For the integration
ambiguity was not  really an ambiguity in the measure \Meas, but rather, an
ambiguity in the region of integration. Here the region of integration (defined
by some cutoff $|t|>\epsilon$) is the same; it is really the measure that
suffers from a potential ambiguity. We believe that there is, in fact, no
ambiguity so long as one restricts oneself to coordinate families with the
correct asymptotic behavior at infinity  ({\it i.e.}~coordinate families which
also give the correct dilaton insertions). A more careful investigation of
this question would be desirable.

What are the implications of our results for calculating string scattering
amplitudes? The simplest scattering amplitude where these issues arise is the
4-point function on the sphere. $\mh 04$ has dimension $1|2$, and the boundary
of $\mh 04$ consists of three divisors of dimension $0|2$. Each boundary
component
corresponds to the sphere pinching in two, with two punctures landing on
either side of the pinch. When all of the punctures are far apart, we can
choose
coordinates for supermoduli space by saying that the four punctures are located
 at $(0,0)\ (1,\rho_1),\ (r,0)$ and $(\infty,\rho_2)$. We also demand that the
local coordinate at each puncture depend only on the moduli associated to that
puncture. Near each of the boundaries of supermoduli space, however, these
coordinates degenerate, and must be replaced by those associated to the
pinching geometry. The transformations between the coordinates
($r,\rho_1,\rho_2$) and those appropriate near the boundary must involve a
non-split transformation for at least one of the components of the boundary.
In the case at hand, it is $r\to 0$ which involves the nonsplit
transformation but, however we choose to coordinatize the interior of
supermoduli space, it is impossible to avoid a nonsplit transformation
somewhere \Meas.
In constructing the corresponding string measure, this nonsplit
transformation manifests itself as a correction to the ``na\" \i ve" measure,
which restores the superconformal Ward identities. The calculation of the
measure in the interior of moduli space is easily translated into the
conventional picture-changing language. The vertex operators at $1,\infty$ are
in the
(0)-picture, while the vertex operators at $0,r$ are in the $(-1)$-picture.
But the picture-changing is too na\" \i ve, and in need of corrections when the
two $(-1)$-picture vertex operators run into each other.

The real point of this is that, in general, we need to cut off the integration
because the kinematics may be such that the
measure diverges near some boundary of moduli space.
If we simply impose a cutoff
$(|r|>\epsilon)$,
in the picture-changing formalism, we obtain a surface term which
not superconformally covariant \EFEFS. To it, Green and Seiberg
were forced to add
a contact term. A better approach is to impose a cutoff on the pinching
parameter ($|t|>\epsilon$). The measure constructed in the operator formalism
is {\it automatically}
superconformally invariant. With this choice of cutoff, no
extra contact term is required. In
particular, the contact term was required to remove the contribution of states
like the bosonic string tachyon which are not even in the spectrum of the
fermionic string theory. In the present formalism, such states are already
projected out. The {\it only} divergences of the string measure in the
present formalism are due to {\it physical} on-shell poles.

\bigbreak\bigskip\bigskip\centerline{{\bf Acknowledgments}}\nobreak
\frenchspacing{
We would like to thank D. Kutasov, P. Nelson and H. Verlinde for useful
conversations. This work was supported by NSF grant PHY90-21984.
}

\listrefs
\end